\documentclass[prb,aps,superscriptaddress,groupedaddress,twocolumn]{revtex4}

\usepackage{xspace,amsmath,amsfonts,amsthm,amsbsy}
\usepackage{graphicx}
\usepackage{amssymb}
\usepackage{color}
\usepackage{epsfig,psfrag,subfigure,subeqnarray}
\usepackage[outdir=./]{epstopdf}
\usepackage[usenames,dvipsnames]{xcolor}
\usepackage[colorlinks=true,citecolor=Blue,linkcolor=RubineRed,urlcolor=Blue]{hyperref}

\usepackage{ulem}
\usepackage{booktabs}

\newcommand{\ra}[1]{\renewcommand{\arraystretch}{#1}}

\definecolor{pink}{rgb}{1,0.078,0.57}

\definecolor{green}{rgb}{0,0.7,0.9}

\newcommand{\ket}[1] {\left\vert #1 \right\rangle}
\newcommand{\bra}[1] {\left\langle #1 \right|}
\newcommand{\braket}[2] {\langle #1 | #2 \rangle}

\newcommand{\bk}{\mathbf{k}}
\newcommand{\bp}{\mathbf{p}}
\newcommand{\bq}{\mathbf{q}}
\newcommand{\bQ}{\mathbf{Q}}
\newcommand{\br}{\mathbf{r}}

\newcommand{\dg}{^{\dagger}}

\newcommand{\ankp}{a_{n_e,\bk_\parallel}}
\newcommand{\bnkp}{b_{n_h,\bk_\parallel}}

\newcommand{\rp}{\br_\parallel}
\newcommand{\kp}{\bk_\parallel}

\newcommand{\Qp}{\bQ_\parallel}

\newcommand{\hc}{\mathrm{h.c.}}

\newcommand{\vac}{\ket{v}}

\begin{document}
\title{From hybrid polariton to dipolariton using \\
non-hermitian Hamiltonians to handle particle lifetimes}

\author{Aur\'elia Chenu}
\affiliation{Department of Physics and Materials Science, University of Luxembourg, 1511 Luxembourg, G.D. Luxembourg}

\author{Shiue-Yuan Shiau} 
\affiliation{Physics Division, National Center for Theoretical Sciences, Taipei 10617, Taiwan}

\author{Ching-Hang Chien} 
\affiliation{Research Center for Applied Sciences, Academia Sinica, Taipei 115, Taiwan}

\author{Monique Combescot}
\affiliation{Sorbonne Universit\'e, CNRS, Institut des NanoSciences de Paris, 75005 Paris, France}

\begin{abstract}
We consider  photons strongly coupled to  the excitonic excitations of a coupled quantum well, in  the presence of an electric field. We show how under a field increase, the hybrid polariton made of  photon  coupled to hybrid carriers lying in the two  wells,  transforms into a dipolariton made of photon  coupled to direct and indirect excitons. We also show how the cavity photon lifetime and  the  coherence time of the carrier wave vectors, that we analytically handle through  non-hermitian Hamiltonians, affect these polaritonic states.  While the hybrid polaritons display a spectral singularity, where the eigenvalues coalesce and known as exceptional point, that depends on  detuning and lifetimes, we find that the three dipolaritonic states display an anti-crossing without  exceptional point, due to interaction between photons, direct and indirect excitons. 
\end{abstract}

\maketitle

\section{Introduction \label{sec1}}

Coupled quantum wells (CQWs) provide a unique platform  to study the rich quantum properties of excitons \cite{Miller1985a, Le1987a,butov1999,larionov2000,butov2002,combescot2017} that  have found numerous applications in optoelectronics \cite{Ahn1989a, Tokuda1990a, High2007a, High2008a}. Of particular interest is a CQW  under  an  electric field perpendicular to the well plane: electrons and holes, which  for high enough field are separated in the two adjacent wells, form spatially indirect excitons, in contrast to direct excitons that are formed from electrons and holes in the same well. 
The charge separation generates a built-in electric dipole moment, that can be controlled by the applied electric field. This has been used to control exciton transport  in  electrostatic \cite{Winbow2011a, schinner2013} or optical \cite{alloing2013, hammack2006} ways, and to develop excitonic devices \cite{andreakou2014}. 
Such indirect excitons have a long lifetime ranging from 10 ns to 10 $\mu$s \cite{High2008a}, which exceeds that of  direct excitons by orders of magnitude due to the small overlap between electron and  hole wave functions  that strongly decreases their  radiative recombination \cite{Kash1985a,Voros2005a, sivalertporn2012,sivalertporn2015, Beian2015a}. Thus, they can easily reach  thermal equilibrium and allow exploring collective quantum phenomena   such as  exotic many-body phases \cite{anankine2017} and exciton Bose-Einstein condensation \cite{Beian2017a}.

\begin{figure*}
\begin{center}
\includegraphics[width=15cm]{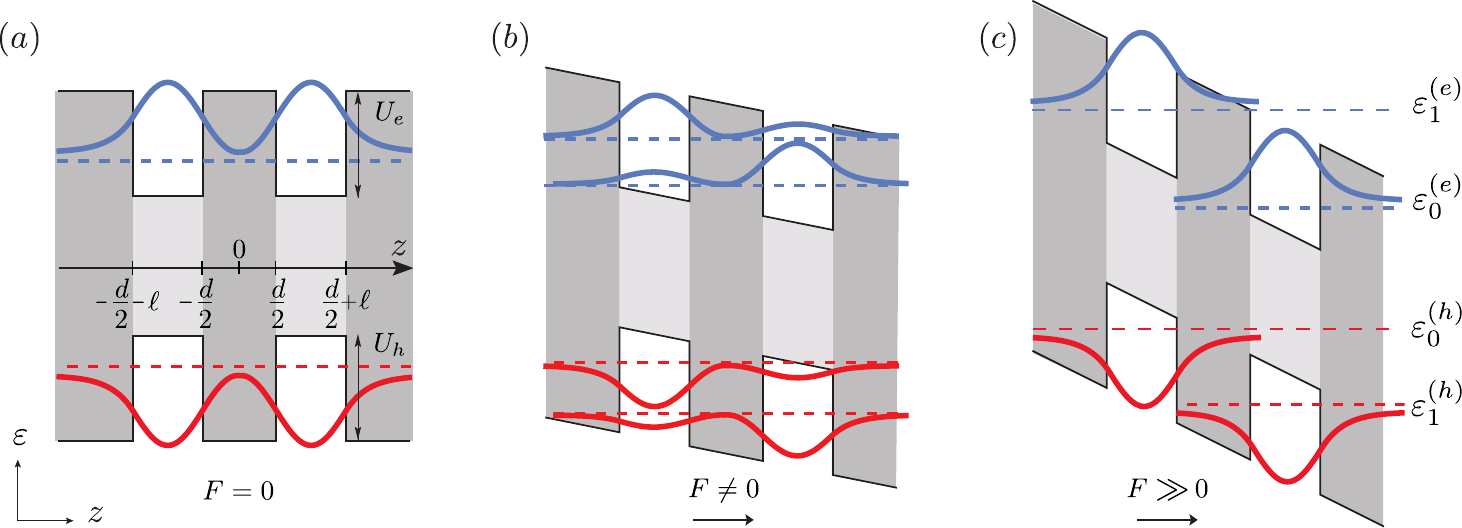}
\caption{Schematic representation of carrier energies (dashed lines) and wave functions (solid lines) for electrons (blue) and holes (red) in a coupled quantum well, without (a) or with (b,c) electric field  $F$. The ground level $\varepsilon_0^{(i)}$ for $F=0$ (a)  transforms into two levels under increasing $F$ (b,c). 
}
\label{fig1}
\end{center}
\end{figure*}

The small overlap between electron and hole wave functions renders indirect excitons  quite long-lived, but it also makes them weakly coupled to light. A way to enhance the photon coupling to excitons is to embed the CQW  in a microcavity \cite{cristofolini2012}. The resulting exciton-photon particles can be transformed from hybrid polaritons to dipolaritons by applying an external field.  
In contrast to hybrid  polariton made from cavity photons strongly coupled to hybrid carriers that belong to the two quantum wells, the dipolaritonic modes result  from the strong  coupling of  indirect excitons,  direct excitons, and  cavity photons. The  controllability of the resulting system has  inspired  new applications such as  quantum logic gates \cite{kyriienko2016}, optical parametric oscillators \cite{khadzhi2015}, tunable single-photon emission \cite{kyriienko2014} and dipolariton Bose-Einstein condensation \cite{shahnazaryan2015}. 

 In this paper, we consider a CQW  in an optical cavity, and we study how the cavity photon lifetime and the carrier coherence time  affect the dipolariton eigenstates. 
 We wish to recall that, for polaritonic modes to be formed, the  coherence time of the exciton wave vector must be long at the exciton-photon coupling scale; otherwise  the exciton wave vector $\bQ$ would change and the exciton would emit a different photon $\bQ' \neq \bQ$. The initial $\bQ$ photon would thus be lost, and the polaritonic mode $\bQ$ cannot develop \cite{dubin2005}. 
 
 We here focus on the  two limiting cases: \\
 (1) In the absence of electric field (see Fig.~\ref{fig1}(a)), the electron and hole states result from the hybridization of individual quantum well states, depending on the  barrier height and well thickness. The excitons formed from these hybridized states are coupled to photons. The resulting polariton branches then have a lifetime that  depends on the photon detuning and the photon and carrier lifetimes. \\
 (2) When the electric field is large (see Fig.~\ref{fig1}(c)), the electron ground state with energy $\varepsilon_0^{(e)}$ is mainly localized in one well, while the hole ground state with energy $\varepsilon_0^{(h)}$ is mainly in the other well. The first excited states of these carriers, with energies $\varepsilon_1^{(e)}$ and  $\varepsilon_1^{(h)}$, are mainly localized the other way around, provided that  the electric field is not too large; otherwise the carrier first excited state could be located in the same well as the ground state \cite{sivalertporn2016}. The  exciton ground state, made of carriers in the $(\varepsilon_0^{(e)}, \varepsilon_0^{(h)})$ levels, is an indirect exciton. The next higher excitonic states, made of carriers in the same well, are direct excitons. The  quantum nature of the particles  allows them  to penetrate  the barriers. This induces a  small but finite overlap between the electron and hole wave functions. 
  The indirect exciton thus has a small but direct coupling to photons, with no need of any tunneling process that could induce a coupling to photons through  direct excitons. 
The fact that the photon is coupled to both direct and indirect excitons leads to  eigenstates,  named dipolaritons, whose lifetime depends not only on the photon detuning to direct and indirect excitons, but also on the lifetimes of  excitons and cavity photons.

The exciton binding energies and the resulting excitonic photoluminescence spectra in a CQW under a static electric field have been studied using  variational methods \cite{Kamizato1989a, Dignam1991a, Soubusta1999a}, or  by numerically solving the Schr\"odinger equation  of the coupled carriers \cite{sivalertporn2012,sivalertporn2015}. We here propose an analytical approach to study these energies, in which the particle lifetimes is handled through the non-hermitian Hamiltonian approach.

 The paper is organized as follows:
 \\
$\bullet$ In section  \ref{sec2}, we explain how to analytically study cavity photons coupled to quantum well excitons without and with a bias voltage between the two wells. Particular attention is paid to the photon coupling with indirect excitons, that was previously understood \cite{cristofolini2012} through  electron tunneling between the wells, instead of by just taking into account the existing leakage of the carrier wave functions. 
We show how to derive the hybrid polariton and dipolariton eigenstates when the particle  lifetimes are finite. These lifetimes    bring imaginary parts to the particle energies, that mathematically lead  to non-hermitian Hamiltonians with different bra and ket eigenstates. The diagonal form of the Hamiltonian then appears in terms of  creation operators for ket eigenstates, and destruction operators for bra eigenstates. Using this non-hermitian formalism, it becomes easy to analytically derive the time evolution of  photons coupled to electronic excitations located in two quantum wells, without and with a bias voltage, and to see the transition from hybrid polariton to dipolariton  when the detuning and particle lifetimes change.
 \\
 $\bullet$ The unbiased case (no field)  is presented in Sec.~\ref{sec3}, with photons coupled to excitons made of electrons and holes in hybridized states. The Hamiltonian then depends on  two operators only, the creation operators for photon and  hybridized excitons. 
 \\
$\bullet$  In  Sec.~\ref{sec4}, we consider a bias voltage large enough so that the direct excitons, made of electrons and holes in the same quantum well, are  well separated in energy from the indirect excitons, made of electrons and holes in different wells. Two kinds of direct excitons and two kinds of indirect excitons \textit{a priori} exist. While the two indirect excitons are well separated in energy so that the one with  the highest energy can be eliminated from the problem, the two direct excitons have close energies; they would be equal for identical carrier leakages. It is possible to show that, for equal coupling  between photon and   direct excitons, one combination of direct excitons is decoupled, while the other gains  a $\sqrt{2}$ bosonic enhancement. This leads to a three-body Hamiltonian made of photon, direct and indirect excitons, that is, a rank-3 eigenvalue equation which can be analytically solved using the Cardan's trick (see Appendix A). \\
 $\bullet$   The main equations derived in this work are summarized in Table \ref{table1}, at the end of the manuscript.
 
 \section{General formalism \label{sec2}}
 \subsection{Semiconductor Hamiltonian}
 We consider two quantum wells having  equal width $\ell$ separated by a potential barrier with thickness $d$, as shown in Fig.~\ref{fig1}. This CQW structure can be made of two layers of InGaAs materials embedded in a GaAs material. Since the GaAs gap is larger than the InGaAs gap, the electron and hole wave functions along the growth axis $z$ tend to localize inside the InGaAs wells. The carriers  move freely in the well planes in the absence of Coulomb interaction.

 The semiconductor Hamiltonian for one electron and one hole reduces to 
 \begin{equation}\label{eq:Hsc}
 H_{eh} =H_e + H_h + V_{eh}\,. 
 \end{equation}
The ($H_e, H_h$) parts consist of  kinetic energies and  quantum well potentials for the electron and the hole, with a   Coulomb interaction $V_{eh}$ between them. 
 
\noindent\textbf{(1) One-body parts}: 
in the presence of external electric field $F$, the Hamiltonian for the carrier $i=(e,h)$  reads, in first quantization, as 
 \begin{equation}\label{eq:He}
 H_i = \frac{\bp^2}{2 m_i} + U_i(z)\mp |e|Fz    \,,
\end{equation}
 where the minus and plus sign in front of the field energy refers  to  electron and  hole, respectively. The potential $U_e(z)$ is equal to the difference between the  conduction band bottoms of the two semiconductors composing the quantum well, with $U_h(z)$  for the valence band  tops. Specifically, 
 \begin{equation}
U_i(z)= -U_i  \,\Theta(z) \,,
\end{equation} 
  with $\Theta(z)=1$ inside the wells,  $d/2 < |z| < \ell+d/2$, and $\Theta(z)=0$  outside (see Fig.~\ref{fig1}).

By writing $\bp^2$ as $p_z^2+\bp_{\parallel}^2$, we can separate the electron Hamiltonian $H_e$ into a  part $\bp_{\parallel}^2/2 m_e$ that describes the free motion in the well plane, and a part  that depends on $z$   
 \begin{equation}
 h_{z,e}=\frac{p_{z}^2}{2 m_e}-U_e  \,\Theta(z) - eFz\,.
 \end{equation}
The wave function of the $H_e$ eigenstate, solution of $\left(H_e-\varepsilon^{(e)}_{n_e,\bk_{\parallel}}\right)|n_e^{(e)},\bk_{\parallel}\rangle=0$, then splits as 
 \begin{equation} 
 \langle \br|n_e^{(e)},\bk_{\parallel}\rangle=\braket{z}{n_e^{(e)}} \, \braket{ \rp}{\kp}=\phi_{n_e}^{(e)}(z) \,\frac{e^{i \kp \cdot \rp}}{L}\,,
  \end{equation}
 where $L^2$ is the well area, in the direction orthogonal to $z$, while $\phi_{n_e}^{(e)}(z)$ is the  wave function of the  $h_{z,e}$ eigenstate $|n_e^{(e)}\rangle$ 
  \begin{equation} 
  0 = \left(  h_{z,e} - \varepsilon_{n_e}^{(e)} \right)|n_e^{(e)}\rangle\,.\label{hzeeigenstate}
  \end{equation}
And similarly for  holes.

 Let $a\dg_{n_e,\kp}$ and $b\dg_{n_h,\kp}$ be the creation operators of the electron and hole states, \begin{subeqnarray}\label{4}
a\dg_{n_e,\kp}\vac &=&|n_e^{(e)},\bk_{\parallel}\rangle \,,\label{4_0}\\
b\dg_{n_h,\kp}\vac &=&|n_h^{(h)},\bk_{\parallel}\rangle\,,\label{4_1}
\end{subeqnarray} 
with $\vac$ denoting the vacuum state. 
 The one-body part of the electron-hole Hamiltonian \eqref{eq:Hsc} then reads in terms of electron and hole operators as  
\begin{eqnarray}
H_e + H_h &\simeq&\sum_{n_e=(0,1)} \sum_{\kp} \left(\varepsilon_{n_e}^{(e)} + \varepsilon_{\kp}^{(e)}\right) \ankp\dg \ankp \label{hehh} \\
&&+ \sum_{n_h=(0,1)} \sum_{\kp} \left( \varepsilon_{n_h}^{(h)} +\varepsilon_{\kp}^{(h)}\right) \bnkp\dg \bnkp\,,\nonumber
\end{eqnarray}
with $\varepsilon_{\kp}^{(e,h)} = \hbar^2 \kp^2/2 m_{e,h}$. 

The $\varepsilon_{n_i}^{(i)}$ energies depend on the electric field $F$. The carrier ground level, $n_i^{(i)}=0$ for $F=0$ transforms under increasing $F$ into a ground and an excited level, denoted as  $|0^{(i)}\rangle$ and $|1^{(i)}\rangle$. For small $F$, the carrier wave functions  are hybridized over the two wells (see Fig.~\ref{fig1}(b)), while for large $F$, they are mainly localized in a single well when the barrier potential is large enough (see Fig.~\ref{fig1}(c)). 
 For small or large $F$, direct electron-hole pairs  have a far larger wave function overlap than indirect pairs, which  makes them  more coupled to photons. As shown below, their large coupling has a significant consequence on the  effective coupling between photons and  indirect pairs. This is why  direct pairs have to be taken into account in the construction of the polaritonic states,  although their energy is higher than for  indirect pairs. By contrast, the other indirect pairs made of electron with energy $\varepsilon_1^{(e)}$  and hole with energy $\varepsilon_1^{(h)}$  have a much higher energy, and thus can be neglected.

\noindent \textbf{(2) Two-body interaction}: 
the Coulomb attraction between one electron and one hole reads  in first quantization as
\begin{equation}
V_{eh} = -\frac{e^2}{\epsilon_{\rm sc}|\br_{e} - \br_{h}|}\,,
\end{equation}
where $\epsilon_{\rm sc}$ is the semiconductor  dielectric constant.
In second quantization, $V_{eh}$ appears in terms of  electron and holes operators  as
\begin{eqnarray}
V_{eh}  &\simeq& - \sum_{n'_e,n'_h}\sum_{n_e,n_h}\sum_{\kp,\kp'} \sum_{\bq_\parallel} \mathcal{V}\Big(_{n'_h,\kp'-\bq_\parallel\:\:\: n_h,\kp'}^{ n'_e,\kp+\bq_\parallel \:\:\: n_e,\kp}\Big)\nonumber\\
&&\times b\dg_{n'_h,\kp'-\bq_\parallel} a\dg_{n'_e, \kp+\bq_\parallel}a_{n_e,\kp}   b_{n_h,\kp'}\,.\label{eq:VcoulV}
\end{eqnarray}
According to the second quantization procedure \cite{CSbook}, the prefactor of this two-body operator is given by 
\begin{eqnarray} \label{eq:vq}
\lefteqn{\mathcal{V}\Big(_{n'_h,\kp'-\bq_\parallel\:\:\: n_h,\kp'}^{ n'_e,\kp+\bq_\parallel \:\:\: n_e,\kp}\Big) = {\int} d\br_e {\int} d\br_h   \frac{e^2}{\epsilon_{\rm sc}|\br_e - \br_h|} } \nonumber \\
&&\times \bigg( \phi_{n'_e}^{(e)}(z_e) \frac{e^{i (\kp + \bq_\parallel)\cdot \br_{e\parallel}}}{L}\bigg)^*\!
 \bigg( \phi_{n'_h}^{(h)}(z_h) \frac{e^{i (\kp' - \bq_\parallel)\cdot \br_{h\parallel}}}{L}\bigg)^* \nonumber \\
&&\times \bigg( \phi_{n_e}^{(e)}(z_e) \frac{e^{i \kp \cdot \br_{e\parallel}}}{L}\bigg) 
  \bigg( \phi_{n_h}^{(h)}(z_h) \frac{e^{i \kp'\cdot \br_{h\parallel}}}{L}\bigg) \,,
\end{eqnarray}
 with  $d\br_i = dz_i \, d\br_{i\parallel}$. After integrating over $(\br_{e\parallel},\br_{h\parallel})$, we find that this scattering does not depend on $(\kp,\kp')$; it  reduces to
\begin{eqnarray}\label{eq:vq_reduced}
\mathcal{V}_{\bq_\parallel}\big(_{n'_h\: n_h}^{n'_e \: n_e}\big)&&=\frac{2\pi e^2}{L^2 \epsilon_{\rm sc}q_\parallel} {\iint} dz_e dz_h   \, e^{-q_\parallel |z_e-z_h|}\\
&&\times \Big(\phi_{n'_e}^{(e)}(z_e) \phi_{n'_h}^{(h)}(z_h)\Big)^* \Big(\phi_{n_e}^{(e)}(z_e)\phi_{n_h}^{(h)}(z_h)\Big)\,,\nonumber
\end{eqnarray}
which evidences its dependence  on the wave function overlap between the  $n_e$ and $n'_e$ electron states and  between the  $n_h$ and $n'_h$ hole states. The largest overlaps correspond to $n_e=n'_e$ and $n'_h=n_h$, that is, to Coulomb processes in which the carriers do not change well, which are the ones leading to direct excitons.

\noindent \textbf{(3) Exciton states}: 
 the last step is to introduce electron-hole pair operators for the physically relevant states. 
The creation operator for a free pair made of an electron in the $n_e$ level and a hole in  the $n_h$ level, with a center-of-mass wave vector $\Qp$ and a relative-motion wave vector $\kp$, reads  
\begin{equation}
B\dg_{n_e,n_h;\Qp,\kp} = a\dg_{n_e, \kp + \mu_e \Qp} b\dg_{n_h, -\kp + \mu_h \Qp}\,,
\end{equation}
with  $\mu_e = 1 - \mu_h = m_e /M_{_X}$ for $M_{_X} = m_e + m_h$.
The electron-hole attraction  transforms this  free pair into a correlated pair, that is, an exciton with a center-of-mass wave vector $\Qp$ and a relative-motion index $\nu$, its creation operator reading as
\begin{equation}\label{corrPair}
B\dg_{\Qp,\nu} = \sum_{n_e,n_h}\sum_{\kp}  B\dg_{n_e,n_h;\Qp,\kp} \braket{n_e,n_h;\kp}{\nu}\,,
\end{equation}
the  exciton energy being equal to  $E_{\Qp,\nu} =  \hbar^2 \Qp^2/2 M_{_X} +\varepsilon_\nu$. 
The $\braket{n_e,n_h;\kp}{\nu}$ prefactor is solution of the  Schr\"{o}dinger equation
\begin{eqnarray}\label{15}
0&=&\left(\varepsilon_{n_e}^{(e)} {+} \varepsilon_{n_h}^{(h)}{+}\frac{\hbar^2 \kp^2}{2\mu_X}{-}\varepsilon_\nu \right)\braket{n_e, n_h;\kp}{\nu}\\
&&-\sum_{n'_e,n'_h}\sum_{\bq_\parallel}\mathcal{V}_{\bq_\parallel}\big(_{n_h\,\: n'_h}^{n_e \,\: n'_e}\big)\braket{n'_e,n'_h;\kp-\bq_\parallel}{\nu},\nonumber
\end{eqnarray}
 where $\mu_{_X}$ is the electron-hole reduced mass given by $\mu_{_X}^{-1}=m_e^{-1}+m_h^{-1}$. The exciton relative-motion energy $ \varepsilon_{\nu}$  depends on the single particle energies, 
 $\varepsilon_{n_e}^{(e)}$ and $\varepsilon_{n_h}^{(h)}$, and the Coulomb scattering $\mathcal{V}_{\bq_\parallel}\big(_{n_h\: n'_h}^{n_e \: n'_e}\big)$;  through them, it also depends on the CQW parameters, $U_e$, $U_h$, $\ell$, and $d$.

%The above correlated pairs, eigenstates of the electron-hole Hamiltonian (\ref{eq:Hsc}), depend on the external electric field.
For problems involving one electron-hole pair only, as the present one, we can replace  the $H_{eh}$ Hamiltonian by its diagonal form 
\begin{equation}\label{Hsc_diagonal}
H_{eh}\simeq\sum_{\Qp}\sum_{\nu}  E_{\Qp,\nu}   B\dg_{\Qp,\nu}  B_{\Qp,\nu} \, ,
\end{equation}
which is convenient to derive polaritons. 

It is important to note that the exciton eigenstates are linear combinations of $n_e$ electrons and $n_h$ holes; the way these  particle wave functions are distributed along $z$  in the two wells depends on the electric field. In general, this distribution  is nontrivial  and can only be numerically obtained. Excitons that are  mainly made of electron-hole pairs located in the same well are called direct, while those mainly made of  electron-hole pairs located in different wells are called indirect. When the carrier wave functions are extended in the two wells, the excitons are called hybrid.

\subsection{Photon-semiconductor coupling}
In this section, we consider a CQW located in a photon cavity. The photon coupling to  direct and indirect excitons is controlled by the electric field through the exciton wave functions. We  first consider photon coupling to a bulk exciton and then turn to a CQW exciton, to better catch the consequences of the well confinement. 

\noindent\textbf{(1) Bulk}: in 3D, a photon with wave vector $\bQ$ and creation operator $\alpha\dg_\bQ$ is coupled to  free pairs with center-of-mass wave vector $\bQ$ and creation operators 
\begin{equation}\label{freeB}
B\dg_{\bQ,\bk} = a\dg_{\bk+ \mu_e \bQ} b\dg_{-\bk + \mu_h \bQ}\,.
\end{equation} 
The photon-semiconductor interaction  then reads  
 \begin{equation}\label{Hph-sc}
 V_{\rm ph-sc} = \sum_\bQ \sum_\bk \Omega_{\bQ, \bk} \,B\dg_{\bQ, \bk}\, \alpha_{\bQ} + \hc \,.
 \end{equation}
The coupling $\Omega_{\bQ, \bk} $ is essentially $\bk$-independent for semiconductors having conduction and valence bands with different parities, as the ones we here consider; so, $\Omega_{\bQ, \bk} \sim \Omega_\bQ \sim \Omega_0$, since the photon wave vector $\bQ$ is very small compared to the relevant electron wave vectors.

 Bulk excitons $ B\dg_{\bQ, \nu}\vac$ are related to free pairs $ B\dg_{\bQ, \bk} \vac$ through 
 \begin{subeqnarray}\label{Bfree}
 B\dg_{\bQ, \nu} &=& \sum_\bk B\dg_{\bQ, \bk} \braket{\bk}{\nu}\,,\\
 B\dg_{\bQ, \bk} &=& \sum_\nu B\dg_{\bQ, \nu} \braket{\nu}{\bk}\,, 
 \end{subeqnarray}
 where $\braket{\bk}{\nu}$ is the bulk exciton relative-motion wave function in momentum space. This gives  
 the $ V_{\rm ph-sc}$ interaction  in terms of excitons as 
 \begin{equation}
 V_{\rm ph-sc} = \Omega_0 \sum_\bQ \sum_\nu  B\dg_{\bQ, \nu} \alpha_{\bQ} \sum_\bk \braket{\nu}{\bk} + \hc\, . 
 \end{equation}
 The  sum over $\bk$ can be calculated by using $\braket{\bk}{\br} = e^{-i \bk \cdot \br}  / L^{3/2}$: if we multiply $\sum_\bk \braket{\nu}{\bk}$ by $1$ written as $L^{3/2} \braket{\bk}{\br=0}$ and use the $\ket \bk$ state closure relation,  the sum over $\bk$  reduces to $L^{3/2}\braket{\nu}{\br=0}$. 
So, the  photon-semiconductor interaction ultimately appears in terms of 3D exciton creation operators  as 
 \begin{equation}
  V_{\rm ph-sc} = \sum_\bQ \sum_\nu \Omega_\nu \, B\dg_{\bQ, \nu} \, \alpha_\bQ + \hc\,,
 \end{equation}
 with $\Omega_\nu = \Omega_0 L^{3/2}\braket{\nu}{\br=0}$. 
This, in particular,  shows that photons have a larger coupling to low 3D exciton states, due to their smaller spatial extension.

\noindent\textbf{(2) Coupled quantum wells}: the quantum well structure breaks the translational invariance of bulk systems in the $z$ direction. 
 We can derive the photon coupling  from Eq.~(\ref{Hph-sc}), by first writing $B\dg_{\bQ, \bk}$ in terms of the electron and hole states relevant for CQW, that are defined in Eq.~(\ref{4}). This is done by writing the free electron creation operator $a\dg_\bk$ in terms of a complete set of  $|n_e^{(e)}\rangle$ states along $z$ that are defined  in Eq.~(\ref{hzeeigenstate}), namely 
  \begin{equation}
 a\dg_\bk \equiv a\dg_{k_z, \kp} = \sum_{n_e} a\dg_{n_e, \kp} \braket{n_e^{(e)}}{k_z} \,.
 \end{equation} 
And similarly for the hole.  The free pair creation operator $ B\dg_{\bQ, \bk}$ given in Eq.~(\ref{freeB}) becomes, in terms of   $(\ankp\dg,\bnkp\dg)$  operators,  
 \begin{eqnarray}\label{23}
  \lefteqn{B\dg_{Q_z,\Qp; k_z,\kp} 
\!\!= a\dg_{k_z + \mu_e Q_z, \kp+\mu_e \Qp} b\dg_{-k_z + \mu_h Q_z, - \kp+ \mu_h \Qp} }\hspace{1.8cm}
  \nonumber \\
 & = & \sum_{n_e,n_h}\braket{n_e^{(e)}}{k_z {+} \mu_e Q_z} \braket{n_h^{(h)}}{{-}k_z {+} \mu_h Q_z} \nonumber\\
 &&\times a\dg_{n_e, \kp + \mu_e \Qp} b\dg_{n_h,-\kp {+} \mu_h \Qp}\,, 
 \end{eqnarray}
 $ \braket{k_z}{n_e^{(e)}}$ being the Fourier transform of the $\braket{z}{n_e^{(e)}}$ electron wave function. 
 In the following, we will  restrict the sum over quantum-well levels to the  two lowest states,  $|0^{(i)}\rangle$ and $|1^{(i)}\rangle$. 

 The next step is to transform the free pair creation operator $a\dg_{n_e, \kp + \mu_e \Qp} b\dg_{n_h,-\kp {+} \mu_h \Qp}$  of Eq.~(\ref{23}) into creation operators for excitons with a center-of-mass wave vector $ \Qp$, as  in Eq.~(\ref{corrPair}). We get 
\begin{eqnarray}\label{eq24}
B\dg_{Q_z,\Qp;k_z, \kp} &&=\hspace{-0.5em}\sum_{n_e,n_h} \braket{n_e^{(e)}}{k_z {+} \mu_e Q_z} \braket{n_h^{(h)}}{{-} k_z {+} \mu_h Q_z}\nonumber \\
&&\times\sum_{\nu} B\dg_{\Qp, \nu} \braket{\nu}{n_e,n_h;\kp}\,.
 \end{eqnarray}

 The photon-semiconductor coupling given in Eq.~(\ref{Hph-sc})  then reads, in terms of CQW excitons, as
 \begin{equation}\label{Hphsc-final}
  V_{\rm ph-sc} =  \sum_{Q_z,\Qp} \sum_{\nu} \sum_{n_e, n_h}  \Omega_{n_e, n_h; \nu} \, B\dg_{\Qp, \nu}\,  \alpha_{Q_z, \Qp}  + \hc\,,
 \end{equation}
 with  the prefactor given by 
 \begin{equation}\label{30}
 \Omega_{n_e, n_h; \nu} {=} \Omega_0  L  \braket{\nu}{n_e, n_h;\rp{=}0} \! \int_L \! dz \braket{n_e^{(e)}}{z} \braket{n_h^{(h)}}{z}\,.
 \end{equation}
 To obtain this prefactor, we first note that $L \braket{\nu}{n_e, n_h;\rp=0}$  is equal to $\sum_{\kp}\braket{\nu}{n_e, n_h;\kp}$, as obtained by using $1 = L\braket{\kp}{\rp=0}$ and the  closure relation for 2D states $|\kp\rangle$. Next,  to obtain  the integral over $z$, we have used the fact that the photon wave vector $Q_z$ is far smaller than the relevant electronic wave vectors; so, the  sum over $k_z$ reduces to  
 \begin{eqnarray}
&& \sum_{k_z} \braket{n_e^{(e)}}{k_z} \braket{n_h^{(h)}}{{-} k_z} = \\
 &&\!\! \int_L dz_{e} \int_L dz_{h} \braket{n_e^{(e)}}{z_{e}} \braket{n_h^{(h)}}{z_{h}} 
  \sum_{k_z} \braket{z_{e}}{k_{z}} \braket{z_{h}}{{-}k_z}. \nonumber
 \end{eqnarray}
 As $\braket{z_{h}}{-k_z}$
 $=\braket{k_z}{z_{h}} $, 
 the sum over $k_z$ is equal to $\braket{z_{e}}{z_{h}}$; so, we end  with
 \begin{equation}
 \sum_{k_z} \braket{n_e^{(e)}}{k_z} \braket{n_h^{(h)}}{- k_z} = \int_L dz  \braket{n_e^{(e)}}{z} \braket{n_h^{(h)}}{z}, 
 \end{equation}
 which   corresponds to the overlap of the  electron and hole wave functions inside the wells.

For indirect excitons, the overlap between the  electron and hole localized in different wells  is very small, but differs from zero  for finite barrier height and thickness, because the carrier wave functions  do penetrate into the  barrier. 
 So, indirect excitons are not fully decoupled from photons. As a result, it is not necessary to invoke a tunneling effect between direct and indirect excitons  to explain the coupling between photon and indirect exciton. This tunneling   actually is  another, less direct, way to say that carriers leak out of the well. 
 
\subsection{Cavity photons}
Due to the carrier confinement in the $z$ direction, bulk photons are poorly coupled to CQW excitons. Confining photons in a cavity selects a set of discrete photon wave vectors along the $z$ direction. This is crucial to possibly form polaritons with CQW excitons. 
Indeed, as seen from the photon-semiconductor coupling given in Eq.~(\ref{Hphsc-final}), a $\bQ = (Q_z, \Qp)$ photon is coupled to a CQW exciton with the same $\Qp$ wave vector, whereas the conservation of the $Q_z$ component  is lost due to the well confinement. 
So, the exciton can recombine into a photon with different $Q_z$ having essentially the same energy, except if the cavity forces the  allowed $Q_z$ components to be different enough in energy. 

Let us call $Q_z^{(c)}$ the wave vector of the  cavity photon having the lowest energy. For a cavity  such that  the other possible photon energies are outside the relevant energy range, we can restrict the photon Hamiltonian to 
\begin{equation}\label{33}
H_{\rm ph} \simeq \sum_{\Qp} \omega_{ \Qp} \alpha\dg_{ \Qp} \alpha_{ \Qp}\,,
\end{equation}
with $\alpha\dg_{ \Qp}\equiv \alpha\dg_{Q_z^{(c)} ,\Qp}$ and  $\omega_{\Qp} = v_{ph} \sqrt{ \Qp^2 + (Q_z^{(c)})^2}$, where $v_{ph}$ is the photon velocity in the material at hand.

\subsection{System Hamiltonian relevant to CQW}
By collecting the above results, we can write down the  system Hamiltonian for cavity photons interacting with carriers in the CQW as   
\begin{equation}
H = H_{\rm ph} + H_{eh} + V_{\rm ph-sc}\,, 
\end{equation}
with the photon Hamiltonian  given in Eq.~(\ref{33}). The other two Hamiltonians depend on the electric field applied to the CQW.

(i) \textbf{In the absence of electric field ($F=0$)}, the  exciton ground state is mainly made of  electron-hole pairs in their ground level, $n_i^{(i)}=0$, with their wave functions evenly distributed in the two wells (Fig. \ref{fig1}(a)). 
 The electron-hole Hamiltonian (\ref{Hsc_diagonal}) then reduces to 
\begin{equation}
H_{eh} \simeq \sum_{\Qp}\varepsilon_{0,0;\Qp} B\dg_{0,0;\Qp} B_{0,0;\Qp}\,,
\end{equation}
with  $\varepsilon_{0,0;\Qp} =  \hbar^2 \Qp^2/2 M_{_X}  + \varepsilon_{\nu_0}$ where $\varepsilon_{\nu_0}$ is the exciton ground-state energy obtained by restricting $(n_e, n_h)$ states in Eq.~(\ref{15}) to $(0,0)$. 
These  excitons are coupled to cavity photons having the same $\Qp$ through
\begin{equation}\label{eq:37}
V_{\rm ph-sc} \simeq \Omega_{0,0} \sum_{\Qp} B\dg_{0,0;\Qp} \alpha_{\Qp} + \hc
\end{equation}
with $\Omega_{0,0} \equiv \Omega_{0,0;\nu_0}$ given in Eq.~(\ref{30}).

The system Hamiltonian then splits as $H = \sum_{\Qp} h_{\Qp}$ with $h_{\Qp}$ given by 
\begin{eqnarray}
h_{\Qp}&=&\omega_{\Qp} \alpha_{\Qp}\dg \alpha_{\Qp} +\varepsilon_{0,0;\Qp} B\dg_{0,0;\Qp} B_{0,0;\Qp} \nonumber\\
&&+ \left( \Omega_{0,0} B\dg_{0,0;\Qp} \alpha_{\Qp} + \hc \right)\, .
\end{eqnarray}
In the following, we will drop the $\Qp$ index to simplify the notation, with $h_{\Qp}$ now written as  
\begin{equation} \label{39_0}
h=\omega \alpha\dg \alpha+  \varepsilon_{0, 0} B\dg_{0, 0 } B_{0, 0}+  \big(\Omega_{0, 0} B\dg_{0, 0 }\alpha + \hc  \big)\,.
\end{equation}

(ii) \textbf{When  the electric field is large}, it is necessary to include both direct and indirect excitons into the problem. The Hamiltonian $H$ also splits as a sum of $h_{\Qp}$ Hamiltonians which reads, if we again drop the $\Qp$ index, as 
\begin{eqnarray} \label{39}
h&=&\omega \alpha\dg \alpha+ \sum_{(n_e,n_h) = (0,1)} \varepsilon_{n_e, n_h} B\dg_{n_e, n_h } B_{n_e, n_h}\nonumber\\
&&+\sum_{(n_e,n_h) = (0,1)} \big(\Omega_{n_e, n_h} B\dg_{n_e, n_h }\alpha + \hc  \big)\,.
\end{eqnarray}

As schematically shown in  Fig.~\ref{fig1}(c),  the energies of the two direct excitons, $\varepsilon_{0,1}$ and $\varepsilon_{1,0}$, are very close, and higher than the  $\varepsilon_{0,0}$ energy of the indirect exciton. By contrast,  the energy $\varepsilon_{1,1}$ of the other  indirect exciton is  much higher than  the ones of the other three species of excitons. This is why in the following, we will drop this $(1,1)$ indirect exciton from the $h$ Hamiltonian. 
Moreover, although the carriers  leak out of the well not exactly in the same way,  we will for simplicity take $\varepsilon_{0,1} \simeq \varepsilon_{1,0}$  and $\Omega_{0,1} \simeq\Omega_{1,0}$, the photon coupling to  direct excitons being much larger than the  $\Omega_{0,0}$ coupling to the indirect exciton having the lowest energy.

\subsection{Finite lifetimes}

 The main purpose of this work is to study the effect of finite lifetimes on the coupled photon-CQW system. A mathematically simple and physically intuitive way to include them into the problem is to add an imaginary part to the exciton energies, that is, to replace $\varepsilon_{n_e, n_h}$ by $\tilde{\varepsilon}_{n_e, n_h}=\varepsilon_{n_e, n_h} - i\gamma_{_X}$,  where $ \gamma_{_X}$ denotes the inverse lifetime induced by non-radiative mechanisms that weakly depend on the well level. Indeed, the physical processes responsible for the lifetime we here consider, better called `coherence time', are the ones that cause a change in the exciton center-of-mass wave vector, these processes being  the same in the two wells. 
 We also include the  lifetime of the cavity photon induced by mirror imperfections, through replacing $\omega$ by $\tilde{\omega}=\omega-i\gamma_{ph}$. 
 
The imaginary parts representing these lifetimes render  the system Hamiltonian non-hermitian. 
In the next sections, we show how this can be analytically handled. 

\section{Zero electric field: Hybrid polariton\label{sec3}}
In the absence of electric field (Fig.~\ref{fig1}(a)),  the relevant excitons are  hybrid excitons with carriers lying in the two wells. Including their lifetimes transforms  the  Hamiltonian \eqref{39_0} into 
 \begin{equation} \label{h:zeroF_lf}
h=\tilde{\omega}\, \alpha\dg \alpha +\tilde{\varepsilon}_{0,0} B_{0,0}\dg B_{0,0} +  \left( \Omega_{0,0}B_{0,0}\dg \alpha  + \hc \right)\, ,
\end{equation} 
with 
\begin{eqnarray} 
\tilde{\omega}&=&\omega-i\gamma_{ph}\,,\label{eq37}\\
\tilde{\varepsilon}_{0,0} &=&  \varepsilon_{0,0}-i\gamma_{_X} \label{eq38}\,. 
\end{eqnarray}

\noindent\textbf{(1) Hybrid polariton state and energy}: 
due to the particle lifetimes, $h$ differs from $ h\dg$; so, these two operators have different eigenstates. 
Moreover, since $h$ is not hermitian, its eigenstates are not orthogonal, and their energies are not necessarily real.

$\bullet$ Let us  look for the $h$ eigenstates 
 \begin{equation} \label{40}
0=\big(h-\mathcal{E}\big)\ket{P} 
\end{equation}
as linear combination of exciton state $\ket{X}=B_{0,0}\dg \ket{v}$ and photon state $\ket{\alpha}=\alpha\dg \ket{v}$, namely $\ket{P}=x\ket{X}+y\ket{\alpha}$.
The  eigenstate equation (\ref{40}) for hybrid polariton made of photon coupled to hybrid exciton lying in the two wells leads to 
\begin{equation} \label{1}
0=\Big[x(\tilde{\varepsilon}_{0,0}-\mathcal{E})+y\Omega_{0,0} \Big]\ket{X}+\Big[y(\tilde{\omega}-\mathcal{E})+x\Omega_{0,0}^*  \Big]\ket{\alpha}\,.
\end{equation} 
Its projection over $\bra{X}$ and $\bra{\alpha}$  gives two coupled equations for $(x,y)$, which have a nonzero solution provided that their determinant is equal to zero. 
For \begin{equation}
\label{40bis}
\mathcal{E} \equiv \mathcal{E}' + \frac{\tilde{\omega} + \tilde{\varepsilon}_{0,0}}{2}\, ,
\end{equation}
this determinant appears as 
\begin{equation}
0 = \begin{vmatrix}
-\tilde{\omega}' - \mathcal{E}' &\Omega_{0,0} \\ 
\Omega_{0,0}^*& \tilde{\omega}' - \mathcal{E}' 
\end{vmatrix}
=\mathcal{E}'^2-   \tilde{\omega}'^2 - |\Omega_{0,0}|^2 \,.\label{matrixE'mathcal}
\end{equation}
 for $\tilde{\omega} - \mathcal{E} \equiv \tilde{\omega}' - \mathcal{E}'$. 
The  hybrid polariton energies $\mathcal{E}$ follow from the two solutions of this equation. Let us call $\mathcal{E}_{_X}$ the solution that goes to $\tilde{\varepsilon}_{0,0}$ and $\mathcal{E}_{ph}$ the solution that goes to $\tilde{\omega}$, when the coupling $\Omega_{0,0}$ goes to zero. The associated (unnormalized) eigenstates are given by
\begin{subeqnarray}\label{43}
\ket{P_{_X}}&=&\ket{X}+\frac{\Omega_{0,0}^*}{\mathcal{E}_{_X}-\tilde{\omega}}\ket{\alpha}\, ,\\
\ket{P_{ph}}&=& \ket{\alpha} + \frac{\Omega_{0,0}}{\mathcal{E}_{ph}-\tilde{\varepsilon}_{0,0}} \ket{X}\, .
\end{subeqnarray}

$\bullet$ Since $h$ is non-hermitian, $h\not= h^\dagger$, the $h^\dagger$ eigenstates, that 
  differ from the $h$ eigenstates, also play a role in the problem. The same derivation  shows that the  $h^\dagger$ eigenstates simply read as the $h$ eigenstates with   $(\mathcal{E}, \tilde{\omega},\tilde{\varepsilon}_{0,0})$ replaced by  $(\mathcal{E}^*, \tilde{\omega}^*,\tilde{\varepsilon}_{0,0}^*)$: they read 
\begin{subeqnarray}\label{ketQrmx1}
\ket{Q_{_X}}&=&\ket{X}+\frac{\Omega_{0,0}^*}{\mathcal{E}^*_{_X}-\tilde{\omega}^*}\ket{\alpha}\, ,\\
\ket{Q_{ph}}&=& \ket{\alpha} + \frac{\Omega_{0,0}}{\mathcal{E}^*_{ph}-\tilde{\varepsilon}_{0,0}^*} \ket{X}\, .
\end{subeqnarray}
Note that  $\braket{Q}{P}\neq 0$ for $(\gamma_{ph},\gamma_{_X}) \neq0$, while   $\ket{Q}=\ket{P}$  for  $(\gamma_{ph},\gamma_{_X}) = 0$, as expected. 

$\bullet$ To go further, we introduce the polariton operators associated with  the $h$ and $h^\dagger$ eigenstates, $\ket{P_{_X}}=P^\dagger_{_X}\ket{v}$ and $\ket{Q_{_X}}=Q^\dagger_{_X}\ket{v}$, and similarly for $P^\dagger_{ph}$ and  $Q^\dagger_{ph}$. Using Eqs.~(\ref{43},\ref{ketQrmx1}), these operators, that read
\begin{eqnarray}\nonumber
\:\:P^\dagger_{_X}{=}B_{0,0}^\dagger{+}\frac{\Omega_{0,0}^*}{\mathcal{E}_{_X}{-}\tilde{\omega}} \alpha^\dagger,
\qquad P_{ph}\dg{=} \alpha\dg{+} \frac{\Omega_{0,0}}{\mathcal{E}_{ph}{-}\tilde{\varepsilon}_{0,0}} B_{0,0}\dg ,\quad\\
\:\: Q^\dagger_{_X}{=}B_{0,0}^\dagger{+}\frac{\Omega_{0,0}^*}{\mathcal{E}^*_{_X}{-}\tilde{\omega}^*}\alpha^\dagger,    
\qquad Q_{ph}\dg {=}  \alpha\dg{+} \frac{\Omega_{0,0}}{\mathcal{E}^*_{ph}{-}\tilde{\varepsilon}_{0,0}^*} B_{0,0}\dg ,\quad \nonumber
\end{eqnarray}
 fulfill the commutation relations 
\begin{equation}
\frac{\big[Q_{_X}, P^\dagger_{_X}\big]_-}{\braket{Q_{_X}}{P_{_X}}}=1=\frac{\big[Q_{ph}, P^\dagger_{ph}\big]_-}{\braket{Q_{ph}}{P_{ph}}}\,, 
%\big[Q_X, P^\dagger_{ph}\big]_-=0=\big[Q_{ph}, P^\dagger_X\big]_-\, .
\end{equation}
the other commutators being equal to zero.
These relations lead us to write the closure relation for hybrid polaritons as
\begin{equation}\label{51}
\mathbb{I}= \frac{\ket{P_{_X}}\bra{Q_{_X}}}{\braket{Q_{_X}}{P_{_X}}}+\frac{\ket{P_{ph}}\bra{Q_{ph}}}{\braket{Q_{ph}}{P_{ph}}}\,. 
\end{equation}
 The diagonal form of the $h$ Hamiltonian then reads 
\begin{equation}
h=\mathcal{E}_{_X}\frac{P^\dagger_{_X} Q_{_X}}{\braket{Q_{_X}}{P_{_X}}}+\mathcal{E}_{ph}\frac{P^\dagger_{ph} Q_{ph}}{\braket{Q_{ph}}{P_{ph}}}\, ,\label{hasoperatorPQ}
\end{equation} 
as necessary  to  satisfy  $0=\big(h-\mathcal{E}_{_X}\big)\ket{P_{_X}}$ and $0=\big(h-\mathcal{E}_{ph}\big)\ket{P_{ph}}$, which is  easy to check. 
 The above equations also lead to
\begin{equation}
\big[h,P^\dagger_{_X}\big]_-=\mathcal{E}_{_X}P^\dagger_{_X}\, ,\quad  \big[h,P^\dagger_{ph}\big]_-=\mathcal{E}_{ph}P^\dagger_{ph}.
\end{equation}

Equations (\ref{51},\ref{hasoperatorPQ}) also  provide a compact form for the time evolution operator  in the $\ket{\alpha}\otimes\ket{X}$ subspace, namely  
\begin{equation}
 e^{-iht}=e^{-i\mathcal{E}_{_X}t}\,\frac{\ket{P_{_X}}\bra{Q_{_X}} }{\braket{Q_{_X}}{P_{_X}}}+e^{-i\mathcal{E}_{ph}t}\,\frac{\ket{P_{ph}}\bra{Q_{ph}} }{\braket{Q_{ph}}{P_{ph}}}\, . 
\end{equation}

\begin{figure}[t]
\centering
\includegraphics[width=8.1cm]{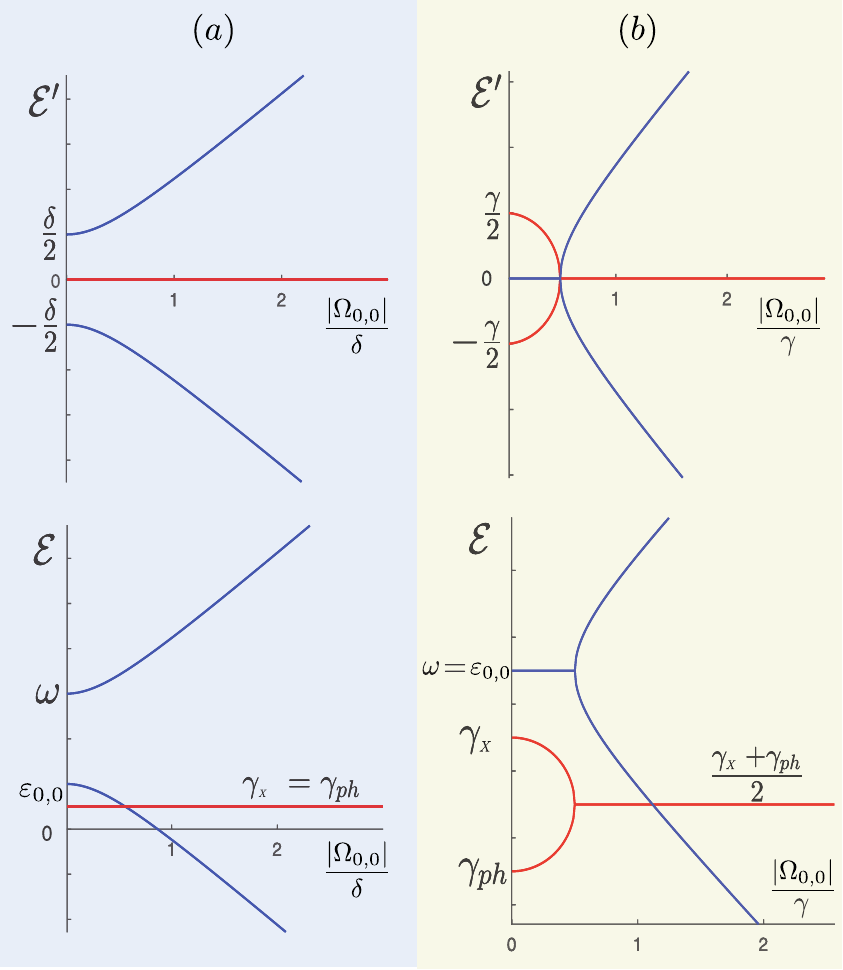}
\caption{
\textbf{Hybrid polariton}. Dependence of the hybrid polariton energies ($\mathcal{E'}, \mathcal{E}$) defined in Eq. (\ref{40bis},\ref{eq59}), on the ratio of the photon coupling to hybrid exciton $\Omega_{0,0}$  (a) over the detuning $\delta$ when the cavity photon lifetime is equal to the exciton coherence time, $0=\gamma=\gamma_{_X}- \gamma_{ph}$, and  (b) over the lifetime difference $\gamma$  at the photon-exciton resonance, $0 = \delta = \omega-\varepsilon_{0,0}$. Real parts (blue) and imaginary parts (red) of ($\mathcal{E'}, \mathcal{E}$). \label{fig2}}
\end{figure}

\noindent\textbf{(2) Hybrid polariton lifetimes}: 
The finite lifetimes of the exciton and cavity photon are inherited by the hybrid polariton through the imaginary part of its eigenvalue given in Eqs. (\ref{40bis}, \ref{matrixE'mathcal}). By writing  
\begin{equation}
\tilde{\omega}-\tilde{\varepsilon}_{0,0}=  \delta + i\gamma
\end{equation} 
where $\delta$ is the photon detuning and $\gamma$ is the  difference between the exciton and photon inverse lifetimes, 
 \begin{equation}
\delta=\omega-\varepsilon_{0,0}, \,\,\,\,\,\qquad  \gamma=\gamma_{_X}-\gamma_{ph}\,, \label{57}
\end{equation}
we can deduce how the photon-exciton interaction couples the two particle lifetimes to the detuning. Let us study this tricky coupling in more detail.

$\bullet$ \textbf{In the absence of photon-exciton interaction} ($\Omega_{0,0}=0$), Eqs. (\ref{40bis}, \ref{matrixE'mathcal}) give $\mathcal{E}'=\pm (\tilde{\omega}-\tilde{\varepsilon}_{0,0})/2$, that is, $\mathcal{E}_{_X} =\tilde{ \varepsilon}_{0,0}$ and $\mathcal{E}_{ph} =\tilde{ \omega}$, as expected. 

$\bullet$  \textbf{When the photon and the exciton have the same lifetime} ($\gamma =0$), the detuning modifies the energies of the two polariton branches, but their lifetimes  stay equal to the common particle lifetime. Indeed, Eqs. (\ref{40bis}, \ref{matrixE'mathcal}) give $\mathcal{E}'=\pm \sqrt{\delta^2 /4+ |\Omega_{0,0}|^2}$ which is  real; so, the two polariton branches have the  lifetime of the particle, regardless of the detuning. 
For a detuning small compared to the photon-exciton coupling,  $\mathcal{E}' \simeq \pm  |\Omega_{0,0}|$, while for large detuning, $\mathcal{E}' \simeq \pm \delta/2$  whatever the coupling, as shown in Fig.~\ref{fig2}(a). 

\begin{figure}[t]
\centering
\includegraphics[width=8.1cm]{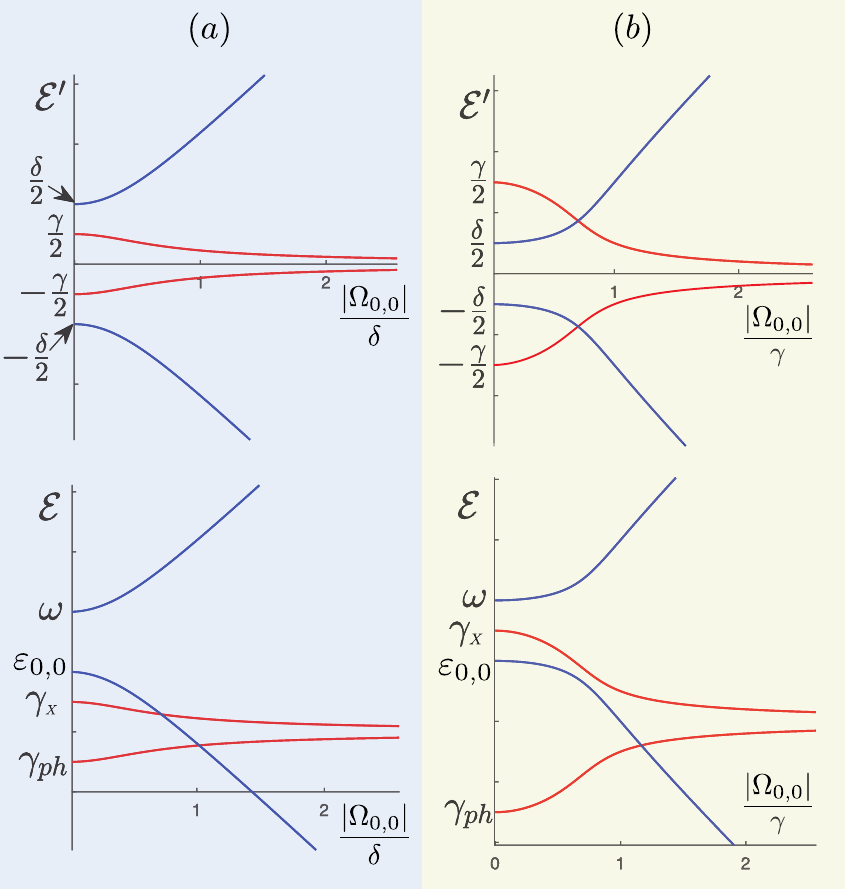}
\caption{\textbf{Hybrid polariton}. Same as Fig.~\ref{fig2},  for  general  $\mathcal{E}'$ and $\mathcal{E}$ given in Eq.~(\ref{eq59}), as a function of the photon-exciton coupling $\Omega_{0,0}$  (a) over the detuning $\delta$ for  $\delta > \gamma>0$  and (b) over the lifetime difference $\gamma$ for $\gamma>\delta>0$. \label{fig3}}
\end{figure}

$\bullet$ \textbf{At the photon-exciton resonance} ($\delta=0$), the energies and lifetimes of the two polariton branches depend on the photon-exciton coupling since  at resonance $\mathcal{E}'$ is equal to $ \pm \sqrt{|\Omega_{0,0} |^2-\gamma^2 /4}$. For small coupling, $|\Omega_{0,0} | <| \gamma|/2$,  the two solutions $\mathcal{E}' = \pm i \sqrt{\gamma^2/4 -  |\Omega_{0,0} |^2}$ are purely imaginary; so, the energies of the two polariton branches  stay equal to the particle energy. Under a coupling increase, their initially different lifetimes converge to the same value when $|\Omega_{0,0}|=| \gamma|/2 $ (see Fig.~\ref{fig2}(b)).  
 This holds until the exceptional point \cite{kato1966,ashida2020},  at which  the two eigenvalues coalesce, while for larger coupling, the coalescence is lifted.
For large coupling, $ |\Omega_{0,0} | > |\gamma|/2 $, the two solutions, $\mathcal{E}' = \pm  \sqrt{  |\Omega_{0,0} |^2-\gamma^2/4}$, are real; so, the two polariton energies depend on the coupling but  their lifetimes stay equal to the lifetime obtained for $|\Omega_{0,0} | = |\gamma|/2 $, that is,
\begin{equation}\label{ave_lifetime}
\bar{\gamma}=\frac{\gamma_{_X}+\gamma_{ph}}{2}\,.
 \end{equation}
The fact that the two polariton branches have the same average lifetime indicates that polaritons  oscillate between photon and exciton faster then the relaxation of any particle.

$\bullet$ These limiting cases help us catch the evolution of the hybrid polariton energies and lifetimes as a function of the photon-exciton coupling. We now consider the general case, that is, photon and exciton having different energies and different lifetimes. To do it, we go back to Eq.~(\ref{matrixE'mathcal}); its two solutions read
\begin{equation} \label{eq59}
\mathcal{E}' = \pm \sqrt{ (\delta + i \gamma)^2/4 + |\Omega_{0,0} |^2 }\,. 
\end{equation}
These two $\mathcal{E}'$ values  evolve from $\pm(\delta  +  i \gamma)/2$ when $|\Omega_{0,0}|= 0$, to $\pm |\Omega_{0,0}|$ for $|\Omega_{0,0}|$ large compared to $|\delta|$ (see Fig.~\ref{fig3}). The energies of the two polariton branches increase with $|\Omega_{0,0}|$, while their lifetimes converge  to the average value given in Eq.~(\ref{ave_lifetime}).
 We note the absence of exceptional point in this setting.
 
All this shows that the coupling between photon and exciton not only changes their energies but also their lifetimes in a tricky way, except when the cavity photon lifetime is equal to the exciton coherence time.

\section{Large electric  field: dipolariton \label{sec4}}

When the external electric field is large (see Fig.~\ref{fig1}(c)), we have shown in Eq.~(\ref{39}) that the relevant Hamiltonian consists of  cavity photon coupled to two direct excitons, $B_{0,1}\dg$ and $B_{1,0}\dg$ with energy $\varepsilon_d\equiv \varepsilon_{0,1}\simeq \varepsilon_{1,0}$, and to  a  ground-state exciton with energy $\varepsilon_{id}\equiv \varepsilon_{0,0}$, which is an indirect exciton, $B_{id}\dg$, its electron and hole being in their ground level localized in different wells. We now add the particle lifetimes. 
  By introducing the two linear combinations of direct excitons, that are normalized and commute,
 \begin{eqnarray}\label{BD_transform}
 B_{d}\dg=\frac{B_{0,1}\dg+B_{1,0}\dg} {\sqrt{2}}  \, ,\quad         D_{d}\dg=\frac{B_{0,1}\dg-B_{1,0}\dg} {\sqrt{2} }\,, 
  \end{eqnarray}
 the $h$ Hamiltonian appears,   after
dropping the (1,1) indirect exciton, as  
 \begin{eqnarray} \label{64}
h&\textcolor{black}{\approx}&\tilde{\omega}\, \alpha\dg \alpha +\tilde{\varepsilon}_{d}\, \Big(B_{d}\dg B_{d}+ D_{d}\dg D_{d}\Big)+\tilde{\varepsilon}_{id}\, B_{id}\dg B_{id}\nonumber  \\ 
 &&+  \Big( \big(\sqrt{2}\, \Omega_{d} \,B_{d}\dg+\Omega_{id} \,B_{id}\dg \big)\alpha + \hc \Big)\, ,
\end{eqnarray}
with the couplings given by $ \Omega_d\equiv \Omega_{0,1}\simeq\Omega_{1,0}$ and  $\Omega_{id}\equiv\Omega_{0,0}$, while the energies are given by  
\begin{eqnarray}
&\tilde{\varepsilon}_d  \equiv\varepsilon_{0,1}-i\gamma_{_X} = \tilde{\varepsilon}_{0,1} \simeq  \tilde{\varepsilon}_{1,0} , \label{eq55} \\
&\tilde{\varepsilon}_{id}\equiv\varepsilon_{0,0}-i\gamma_{_X}.  \label{eq56}
\end{eqnarray}
Note the $ \sqrt{2} $ enhancement factor that appears in the   photon coupling to the $B_d\dg$ direct exciton combination. We also note that  the $D_d\dg$ exciton is not coupled to the cavity photon; so, in the following, we will drop it from the $h$ Hamiltonian given in  Eq. \eqref{64}.

\noindent\textbf{(1) Dipolariton eigenstates}

$\bullet$ 
We proceed as we did for zero electric field. We first look for the $h$ eigenstates, 
$
0=\big(h-\mathcal{E}\big)\ket{P} 
$, as a linear combination of  indirect exciton $\ket{\textsc{ix}}=\,B_{id}\dg \ket{v}$, direct exciton $\ket{\textsc{dx}}=\,B_{d}\dg \ket{v}$ and  cavity photon $\ket{\alpha}=\,\alpha\dg \ket{v}$, namely 
\begin{eqnarray} 
\ket{P}=x_{id}\ket{\textsc{ix}}+x_{d}\ket{\textsc{dx}}+y\ket{\alpha}\,. \label{EQ57}
\end{eqnarray} 
The equation for dipolariton eigenstates then reads 
\begin{eqnarray} 
0&=&\Big(x_{id}(\tilde{\varepsilon}_{id}{-}\mathcal{E}){+}y\Omega_{id} \Big)\ket{\textsc{ix}} \nonumber \\
&+&\Big(x_{d}(\tilde{\varepsilon}_{d}{-}\mathcal{E}){+}\sqrt{2}y\Omega_{d} \Big)\ket{\textsc{dx}}
\\ 
&+&\Big(y(\tilde{\omega}-\mathcal{E})+x_{id}\Omega_{id}^* +\sqrt{2}x_{d}\Omega_{d}^* \Big)\ket{\alpha} \,. \nonumber\label{64_1}
\end{eqnarray} 
Its projection over $\bra{\textsc{dx}}$, $\bra{\textsc{ix}}$ and $\bra{\alpha}$  gives three coupled equations for $(x_{id},x_{d},y)$, which have a nonzero solution provided that their determinant is equal to zero. In terms of 
\begin{equation}
\label{eq58}
\mathcal{E} \equiv \mathcal{E}' + \frac{\tilde{\omega} + \tilde{\varepsilon}_{id}+ \tilde{\varepsilon}_d}{3}\, ,
\end{equation}
 this determinant appears as
\begin{eqnarray}
0 = \begin{vmatrix}
\tilde{\omega}'_{id} - \mathcal{E}' & 0 & \Omega_{id}\\ 
0 &\tilde{\omega}'_{d} - \mathcal{E}'  & \sqrt{2}\,\Omega_{d} \\ 
\Omega_{id}^* &\sqrt{2}\, \Omega_{d}^* & \tilde{\omega}' - \mathcal{E}'
\end{vmatrix}\,,\label{65}
\end{eqnarray}
with  $\tilde{\omega} - \mathcal{E} \equiv \tilde{\omega}' - \mathcal{E}'$; and similarly for $( \tilde{\omega}'_{id}, \tilde{\omega}'_{d})$. 

Let us call $\mathcal{E}_{ph}$ the eigenvalue that tends to $\tilde{\omega}$  when the photon-exciton couplings go to zero, and similarly for ($\mathcal{E}_{id},\mathcal{E}_{d}$). The associated (unnormalized) eigenstates are given by
\begin{eqnarray}
&&\ket{P_{id}}=\ket{\textsc{ix}} \label{ketPrmx1}\\
&&+\frac{\Omega^*_{id}}{(\mathcal{E}_{id}{-}\tilde{\varepsilon}_{d})(\mathcal{E}_{id}{-}\tilde{\omega}){-}2 |\Omega_d|^2}  \left[\sqrt{2}\Omega_d\ket{\textsc{dx}}{+}(\mathcal{E}_{id}{-}\tilde{\varepsilon}_{d})\ket{\alpha}\right]  \nonumber
\end{eqnarray} 
for the indirect-exciton-like branch,
\begin{eqnarray}\label{67}
&&\ket{P_{d}}=\ket{\textsc{dx}} \\
&&+\frac{\sqrt{2}\Omega^*_{d}}{(\mathcal{E}_{d}{-}\tilde{\varepsilon}_{id})(\mathcal{E}_{d}{-}\tilde{\omega}){-}|\Omega_{id}|^2}
 \left[\Omega_{id}\ket{\textsc{ix}}{+}(\mathcal{E}_{d}{-}\tilde{\varepsilon}_{id}\big)\ket{\alpha}\right], \nonumber
\end{eqnarray}
for the direct-exciton-like branch, and 
\begin{equation} \label{Pph}
\ket{P_{ph}}= \ket{\alpha}+
\frac{\Omega_{id}}{\mathcal{E}_{ph}-\tilde{\varepsilon}_{id}} \ket{\textsc{ix}} + \frac{\sqrt{2}\Omega_{d}}{\mathcal{E}_{ph}-\tilde{\varepsilon}_{d}} \ket{\textsc{dx}}\,
\end{equation}
for the photon-like branch.

To get  the $h^\dagger$ eigenstates, $
0=(h^\dagger- \mathcal{E}^*)\ket{Q}$, we again have to replace $\mathcal{E}$ by $\mathcal{E}^*$,  the $\ket{Q}$ eigenstates  being obtained from the $\ket{P}$ eigenstates by changing  $(\tilde{\omega},\tilde{\varepsilon}_{d},\tilde{\varepsilon}_{id})$ into  $(\tilde{\omega}^*,\tilde{\varepsilon}_{d}^*,\tilde{\varepsilon}_{id}^*)$.

$\bullet$ 
To go further, we introduce the creation operators for the $h$ and $h^\dagger$ eigenstates $\ket{P}=P^\dagger \ket{v}$ and $\ket{Q}=Q^\dagger \ket{v}$. These operators fulfill
\begin{equation}
1=\frac{\big[Q_{d}, P^\dagger_{d}\big]_-}{\braket{Q_{d}}{P_{d}}}=\frac{\big[Q_{id}, P^\dagger_{id}\big]_-}{\braket{Q_{id}}{P_{id}}}=\frac{\big[Q_{ph}, P^\dagger_{ph}\big]_-}{\braket{Q_{ph}}{P_{ph}}}\,,
\end{equation}
the other commutators being equal to zero. These relations lead to the following closure relation
\begin{equation}\label{clos_dip}
\mathbb{I}= \frac{\ket{P_{d}}\bra{Q_{d}}}{\braket{Q_{d}}{P_{d}}}+\frac{\ket{P_{id}}\bra{Q_{id}}}{\braket{Q_{id}}{P_{id}}}+\frac{\ket{P_{ph}}\bra{Q_{ph}}}{\braket{Q_{ph}}{P_{ph}}}\,.
\end{equation}

The $h$ Hamiltonian takes a diagonal form in terms of these operators,
\begin{equation}
h=\mathcal{E}_{d}\frac{P^\dagger_{d} Q_{d}}{\braket{Q_{d}}{P_{d}}}+\mathcal{E}_{id}\frac{P^\dagger_{id} Q_{id}}{\braket{Q_{id}}{P_{id}}}+\mathcal{E}_{ph}\frac{P^\dagger_{ph} Q_{ph}}{\braket{Q_{ph}}{P_{ph}}}\, ,\label{69}
\end{equation}
that satisfies $0=(h-\mathcal{E}_{d})\ket{P_d}$,  $0=(h-\mathcal{E}_{id})\ket{P_{id}}$, and  $0=(h-\mathcal{E}_{ph})\ket{P_{ph}}$, as easy to check. Equations (\ref{clos_dip},\ref{69}) also give the time evolution operator  for one photon coupled to  direct and  indirect excitons in a compact form as   
\begin{eqnarray}
e^{-iht}&=&e^{-i\mathcal{E}_{d}t}\frac{|P_{d} \rangle \bra{Q_{d}}}{\braket{Q_{d}}{P_{d}}}+e^{-i\mathcal{E}_{id}t}\frac{|P_{id}\rangle \bra{Q_{id}}}{\braket{Q_{id}}{P_{id}}} \nonumber \\
&+& e^{-i\mathcal{E}_{ph}t}\frac{|P_{ph} \rangle\bra{Q_{ph}}}{\braket{Q_{ph}}{P_{ph}}}\,.\label{70}
\end{eqnarray}

\noindent \textbf{(2) Dipolariton energies and lifetimes}: The eigenvalue equation that follows from the  determinant of Eq.~(\ref{65}) reads
\begin{equation}
 0 = \mathcal{E}'^3 - z_1 \mathcal{E}'^2 + z_2 \mathcal{E}' - z_3 \,,\label{66}
\end{equation}
with $z_i$ given by
\begin{eqnarray}
z_1 &=& \tilde{\omega}'_{id}+\tilde{\omega}'_{d}+\tilde{\omega}' = 0  \,,\nonumber \\
z_2 &=&\tilde{\omega}'_{id}\tilde{\omega}'_{d}+\tilde{\omega}'_{d}\tilde{\omega}'+\tilde{\omega}'\tilde{\omega}'_{id}
-  | \Omega_{id}|^2- 2| \Omega_{d}|^2\,, \label{z2}
 \nonumber \\
z_3 &=&\tilde{\omega}'_{id}\tilde{\omega}'_{d}\tilde{\omega}'- 2| \Omega_d|^2 \tilde{\omega}'_{id}- | \Omega_{id}|^2   \tilde{\omega}'_{d}\, .  \label{z3}
\end{eqnarray}
The  solution of this third-order equation can be analytically obtained by using the Cardan's trick \cite{nickalls1993}  (see Appendix \ref{app:A}). The three solutions read, since $z_1=0$, 
\begin{equation}\label{mathE'n}
\mathcal{E}^{'(n)}=\sum_{s=\pm} e^{i s  \varphi_n }\left[ \frac{z_3}{2} + s\sqrt{ \left( \frac{z_3}{2} \right)^2 + \left( \frac{z_2}{3} \right)^3}\right]^{1/3}  \,,
\end{equation}
with $\varphi_n=(0,\pm2\pi/3)$.

\begin{figure}
\includegraphics[width=8.1cm]{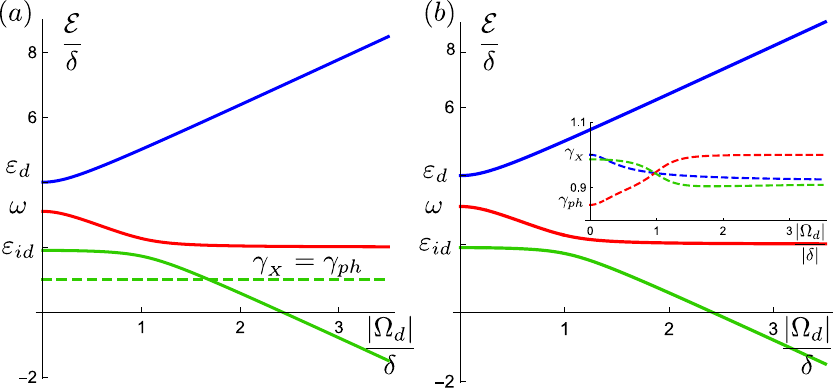}
\vspace{-0.1cm}
\caption{\textbf{Dipolariton}. Dependence of the dipolariton energies $\mathcal{E}$  on the ratio of the photon coupling to \textit{direct} exciton $\Omega_d$ over the detuning with respect to the indirect exciton, $\delta= \omega - \varepsilon_{id}$, when the cavity photon and the exciton have (a) the same lifetime, $\gamma=0$, and (b) a  lifetime difference $\gamma=\delta/6$. 
The other parameters are taken as $|\Omega_{id}|=\delta/3$, $\varepsilon_{id}=2\delta$, $\gamma_{_X}=\delta$, and $\delta_{_X}=\varepsilon_d - \varepsilon_{id}= 2\delta$. The dipolariton has three energy branches given in Eqs.~(\ref{eq58},\ref{z3},\ref{mathE'n}). Their energy real parts are shown as solid lines, while  their inverse lifetimes are shown as dashed lines (inset in (b)). \label{fig4}
}
\end{figure}

\noindent\textbf{(3) Physical understanding}: Let us now discuss how the couplings of the direct and indirect excitons to the \textit{same} cavity photon modify the particle energies and lifetimes; in particular, how a strong photon coupling to the direct exciton modifies the photon coupling to the indirect exciton  when 
\begin{equation}
\lambda\equiv\left| \frac{\Omega_{id}}{ \sqrt{2}\, \Omega_{d}}\right|<1
\end{equation}
as obtained for a  poor overlap between the carriers making the indirect exciton. 

In addition to the ratio $\lambda$ of the photon couplings to indirect and direct excitons, the other two relevant parameters of the dipolariton problem are the photon detuning $\delta$ relative to the indirect exciton ground state and the energy difference $\delta_{_X}$  between direct and indirect excitons
\begin{equation} \label{eq72}
\delta=\omega-\varepsilon_{id}\,,\,\,\,\,\qquad  \delta_{_X}=\varepsilon_{d}-\varepsilon_{id}\,.                                
% \gamma=\gamma_X-\gamma_{ph}\,
\end{equation}
These three  parameters can be experimentally controlled either directly or indirectly. 
The $\tilde{\omega}'$ parameters defined in Eq.~(\ref{65}) then read, for $\gamma=\gamma_{_X} - \gamma_{ph}$ as in Eq.~(\ref{57}),
\begin{eqnarray}
\tilde{\omega}'_{id}&=&(-\delta_{_X}- \delta  - i\gamma)/3 \,,  \nonumber \\
\tilde{\omega}'_{d} &=& (2\delta_{_X}- \delta  - i\gamma)/3 \,,\label{78}\\
\tilde{\omega}' &=& (2\delta- \delta_{_X} + 2i\gamma)/3 \,. \nonumber 
\end{eqnarray}

\begin{figure}
\includegraphics[width=8.1cm]{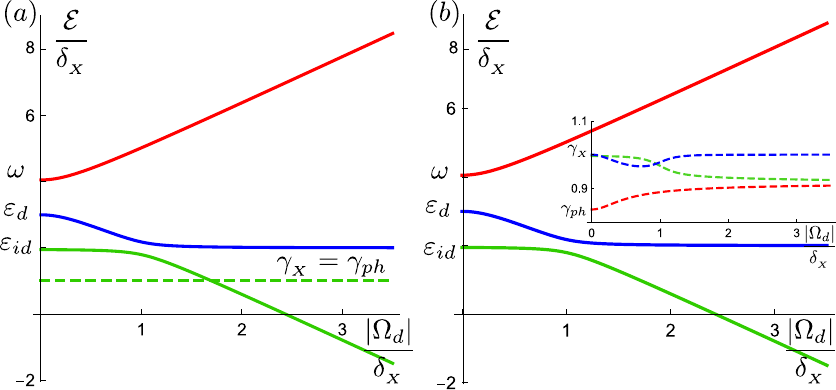}
\vspace{-0.1cm}
\caption{\textbf{Dipolariton}. Same as Fig. \ref{fig4} with ($\delta, \delta_{_X}$) exchanged.  
\label{fig5}}
\end{figure}

\begin{figure*}
\includegraphics[width=\textwidth]{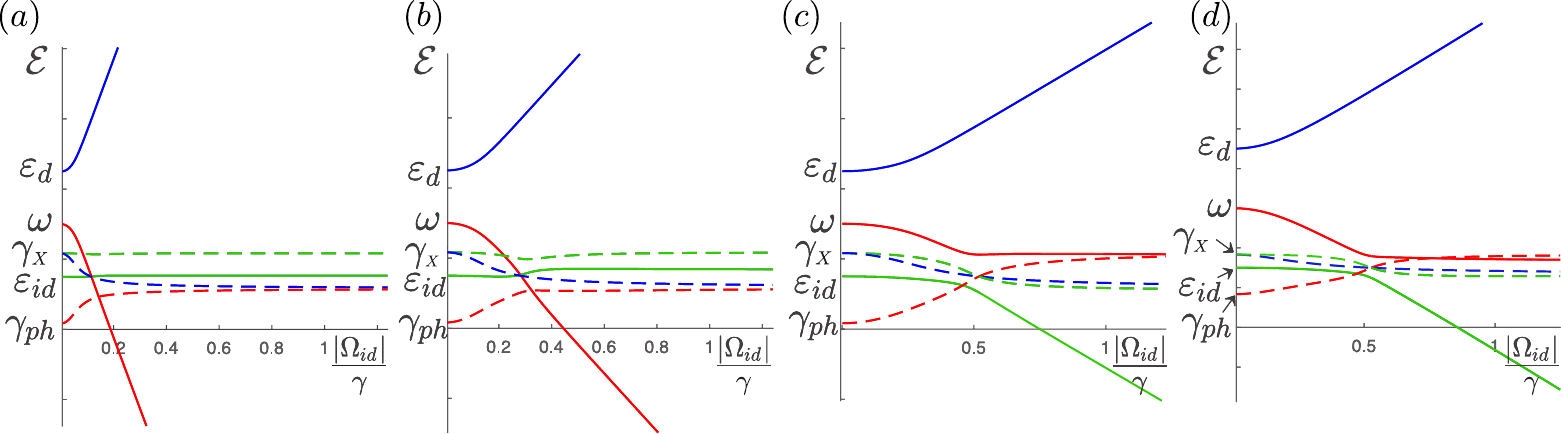}
\caption{\textbf{Dipolariton}. Dependence of the dipolariton energies $\mathcal{E}$ on the ratio of the photon coupling to \textit{indirect} exciton  $|\Omega_{id}|$ over the lifetime difference $\gamma = \gamma_{_X} - \gamma_{ph}$
for various $\gamma$ and coupling ratios $\lambda = |\Omega_{id} / \sqrt{2}\, \Omega_d|$. In (a,b,c), $\gamma$ stays equal to $4 \delta/3$ while $\lambda$ is equal to (0.1; 0.25; 0.5). In (d), $\lambda$ is equal to 0.25 while $\gamma=2\delta/3$. All these curves are calculated for $\delta_{_X} = 2 \delta$. The dipolariton energy real parts are shown as solid lines, while the dashed lines represent the  dipolariton inverse lifetimes. \label{fig6}}
\end{figure*}

\noindent $\bullet$ When  the exciton and photon lifetimes are equal, $\gamma=0$, and when $0 < \delta \ll \delta_{_X}$, that is, when the energy of the direct exciton is much higher than those of the photon and the indirect exciton, the large coupling between photon and direct exciton coupling, $\Omega_d$, makes the upper branch of the dipolariton go up to $\mathcal{E}_{d}\simeq\varepsilon_d+\sqrt{2}\,|\Omega_d|$, and the middle branch go down to $\mathcal{E}_{ph}\simeq \omega-\sqrt{2}\,|\Omega_d|$, until this middle branch anticrosses with the lower branch; the minimum energy splitting between the two is $\simeq 2|\Omega_{id}|$, as  obtained by minimizing the energy difference of the two branches with respect to $|\Omega_d|$. Such an anticrossing exists because the lifetimes of the middle and lower branches are equal. These behaviors are shown in Fig.~\ref{fig4}(a)  for $\delta_{_X}= 2\delta>0$ and $|\Omega_{id}|=\delta/3$: 
 as  $|\Omega_d|$ increases, an anti-crossing occurs between the middle and lower branches, while the middle-branch energy remains flat due to its weak dependence on $|\Omega_{d}|$. So, we end with a hybrid polariton, with the middle branch coupled to the lower branch.

\noindent $\bullet$ When the exciton and photon lifetimes are equal, $\gamma=0$, and when $0<\delta_{_X}\ll \delta$, that is, when the photon energy is much higher than the energies of direct and indirect excitons, the large photon-direct exciton coupling $\Omega_d$ makes the upper branch go up to $\mathcal{E}_{ph}\simeq\omega+\sqrt{2}\,|\Omega_d|$, and the middle branch go down to $\mathcal{E}_{d}\simeq\varepsilon_d-\sqrt{2}\,|\Omega_d|$, until it anticrosses with the lower branch, the minimum energy splitting being  
\begin{equation}
\simeq \frac{2|\Omega_{id}|\sqrt{2|\Omega_{d}|^2+|\Omega_{id}|^2}}{\delta}\,,
\end{equation}
as obtained by minimizing the energy difference of the two branches with respect to $|\Omega_d|$. So, we again end with a hybrid polariton with the middle branch coupled to the lower branch, but the effective coupling between the two branches is different from the one for the $0<\delta\ll \delta_{_X}$ case. The minimum splitting at the anticrossing shown in  Fig.~\ref{fig5}(a), is  smaller than $2|\Omega_{id}|$ as seen from Fig.~\ref{fig4}(a).

\noindent $\bullet$ When the exciton and photon have different lifetimes, $\gamma=\gamma_{_X} - \gamma_{ph}\not=0$,   the  lifetimes of the three dipolariton branches depend on the photon-direct exciton coupling $|\Omega_d|$, as seen from the differences between Figs.~\ref{fig4}(a)-\ref{fig5}(a) and Figs.~\ref{fig4}(b)-\ref{fig5}(b). Surprisingly, for $\gamma=|\Omega_{id}|/2\neq0$,  the inverse lifetimes of the three dipolariton branches go to a single value as $|\Omega_{d}|/\delta$ increases (see Fig.~\ref{fig4}(b)), while the middle and lower branches anticross with a splitting smaller than the one for $\gamma=0$. 
Before this point, the photonic branch has the largest lifetime, while  for larger $|\Omega_d|$, this branch has the smallest lifetime,  the two inverse excitonic lifetimes converging to a smaller value in the large $|\Omega_{d}|$ limit. 
Indeed, for $\delta \ll \delta_{_X}$, the splitting at the anticrossing is   approximately equal to $2\sqrt{(1+\delta/\delta_{_X})|\Omega_{id}|^2-\gamma^2/4}$. 
Equal inverse lifetimes mathematically mean that the three $\mathcal{E}'$ solutions of Eq.~(\ref{66})  are real. This provides a tool to mitigate the short non-radiative lifetime of the excitons. 

When the upper branch is photon-like, as shown in Fig.~\ref{fig5}(b), the middle and lower exciton-like branches anticross; their inverse lifetimes  go  to a single value at $|\Omega_d| = \delta_{_X}$, while the inverse lifetime of the upper branch, that remains the smallest for all $|\Omega_{d}|$ values, goes to the lower branch value for large $|\Omega_{d}|$.

\noindent $\bullet$ We now turn to the effect of the photon couplings to indirect and direct excitons, through their ratio $\lambda$. When $\lambda$ is small, see Figs.~\ref{fig6}(a,b), the energies of the middle and lower branches can cross because the lifetimes of these two branches are different whatever  $|\Omega_{id}|$. The energy and lifetime of the branch that start from the indirect exciton (green curves) remain flat as $|\Omega_{id}|$ increases, which is  the signature of  indirect exciton being weakly coupled to  photon and to direct exciton.  This is in contrast to Fig.~\ref{fig4}(b) which exhibits an anticrossing: as  $\lambda$ increases, the lower two branches anticross and the lifetimes of the three branches go to the same value, as shown in Fig.~\ref{fig6}(c) for  $\lambda=0.5$, which is reminiscent of the results in Fig.~\ref{fig4}(b). 
Note that we can also obtain an anticrossing by decreasing the inverse lifetime difference $\gamma$, as can be seen by comparing Fig.~\ref{fig6}(b) and Fig.~\ref{fig6}(d).

\begin{table*}\centering
\ra{1.4}
\begin{tabular}{@{}l c c c  cc@{}}
\toprule[1.1pt]
&$F=0$ &Eqs.  &\phantom{a}&$F\gg 0$ &Eqs. \\
\cmidrule[0.5pt]{2-3} \cmidrule[0.5pt]{5-6}
Hamiltonian  $h$ &  
$h=\tilde{\omega}\, \alpha\dg \alpha +\tilde{\varepsilon}_{0,0} B_{0,0}\dg B_{0,0}$&& &
$h\approx\tilde{\omega}\, \alpha\dg \alpha +\tilde{\varepsilon}_{d}\, \big(B_{d}\dg B_{d}+ D_{d}\dg D_{d}\big)+\tilde{\varepsilon}_{id}\, B_{id}\dg B_{id}$&\\ 
&  +  $\big( \Omega_{0,0}B_{0,0}\dg \alpha  + \hc \big)$&\eqref{h:zeroF_lf} && $+  \Big( \big(\sqrt{2}\, \Omega_{d} \,B_{d}\dg+\Omega_{id} \,B_{id}\dg \big)\alpha + \hc \Big)$ &\eqref{64}\\
%\hline
Eigenvectors: &&&&\\
 $(h - \mathcal{E}) \ket{P}=0$ &  $\ket{P_{_X}},\: \ket{P_{ph}}$ &\eqref{43} &
 & 
 $\ket{P_{id}}, \: \ket{P_{d}},\: \ket{P_{ph}}$ & (\ref{ketPrmx1}-\ref{Pph})\\ 
$(h\dg - \mathcal{E}^* )\ket{Q}=0$ &  $\ket{Q_{_X}},\: \ket{Q_{ph}}$ &\eqref{ketQrmx1}& &
$\ket{Q_{id}}, \: \ket{Q_{d}},\: \ket{Q_{ph}}$ \\ 
&&&&\\
Diagonal form& 
 $h=\mathcal{E}_{_X}\frac{P^\dagger_{_X} Q_{_X}}{\braket{Q_{_X}}{P_{_X}}}+\mathcal{E}_{ph}\frac{P^\dagger_{ph} Q_{ph}}{\braket{Q_{ph}}{P_{ph}}}$&\eqref{hasoperatorPQ} && 
 $h=\mathcal{E}_{d}\frac{P^\dagger_{d} Q_{d}}{\braket{Q_{d}}{P_{d}}}+\mathcal{E}_{id}\frac{P^\dagger_{id} Q_{id}}{\braket{Q_{id}}{P_{id}}}+\mathcal{E}_{ph}\frac{P^\dagger_{ph} Q_{ph}}{\braket{Q_{ph}}{P_{ph}}}$ &\eqref{69}\\ 
&&&&\\
%\hline
Eigenenergies $\mathcal{E}$ &  $\mathcal{E} \equiv \mathcal{E}' + \frac{\tilde{\omega} + \tilde{\varepsilon}_{0,0}}{2}$ &\eqref{40bis}&
& $\mathcal{E} \equiv \mathcal{E}' + \frac{\tilde{\omega} + \tilde{\varepsilon}_{id}+ \tilde{\varepsilon}_d}{3}$& \eqref{eq58}\\ 
with $\mathcal{E}'$ &  $\mathcal{E}' = \pm \sqrt{ (\delta + i \gamma)^2/4 + |\Omega_{0,0} |^2 }$& \eqref{eq59} && $
\mathcal{E}^{'(n)}=\sum_{s=\pm} e^{i s  \varphi_n }\left[ \frac{z_3}{2} + s\sqrt{ \left( \frac{z_3}{2} \right)^2 + \left( \frac{z_2}{3} \right)^3}\right]^{1/3} $ &\eqref{mathE'n}\\ 
%\hline
Parameters&$\tilde{\omega} = \omega - i \gamma_{ph}; \quad \tilde{\varepsilon}_{0,0} = \varepsilon_{0,0} - i \gamma_{_X}$&(\ref{eq37}-\ref{eq38})&
&$\tilde{\omega} = \omega - i \gamma_{ph}; \quad \tilde{\varepsilon}_{d} = \varepsilon_{0,1} - i \gamma_{_X}; \quad \tilde{\varepsilon}_{id} = \varepsilon_{0,0} - i \gamma_{_X}$&(\ref{eq55}-\ref{eq56})\\
&$\delta = \omega - \varepsilon_{0,0}; \quad \gamma = \gamma_{_X} - \gamma_{ph}$&\eqref{57}&
&$\delta = \omega - \varepsilon_{id};\quad \delta_{_X} - \varepsilon_d  - \varepsilon_{id}; \quad  \gamma = \gamma_{_X} - \gamma_{ph}$& \eqref{eq72}\\
&$\tilde{\omega}' = \tilde{\omega} - (\tilde{\omega}  + \tilde{\varepsilon}_{0,0})/2$&&
&$\{\tilde{\omega}'_{id}, \tilde{\omega}'_d, \tilde{\omega}'\} = \{\tilde{\varepsilon}_{id}, \tilde{\varepsilon}_d, \tilde{\omega}\} - (\tilde{\omega} + \tilde{\varepsilon}_{id} + \tilde{\varepsilon}_d)/3 $&\eqref{78}\\
\bottomrule[1.1pt]
\end{tabular}
\caption{Summary of the main equations and the parameters used in the manuscript. \label{table1} }
\end{table*}

\section{Conclusion}
In this work, we studied how cavity photons couple to direct and indirect excitons in a coupled quantum well when the external electric field increases, with a particular focus on the effect of the particle lifetimes. 
To do so, we first constructed the relevant system Hamiltonians step by step, from scratch. Their diagonal form allowed us to identify the various states that are relevant to the problem at hand. 
The procedure to derive diagonal Hamiltonian operators can be easily extended to multilevel systems. 
Then, we showed how, at low electric field, the photon couples to a hybrid exciton made of carriers lying in the two quantum wells, to form a hybrid polariton. 
By contrast, at large electric field, the photon couples to one indirect exciton and two direct excitons, then forming a dipolariton. 
By adding the lifetime of the cavity photon and the coherence time of the exciton carriers, we have derived the consequences of the carrier relaxation processes and  precisely tracked  the time evolution of this open quantum system. 
We have been able to do it  with the help of non-hermitian Hamiltonians. 
By showing how the polariton energies are affected by these lifetimes, our work  provides physical insights 
 to possible identify the parameter regime in which dipolariton can be created.

\begin{acknowledgements}
We wish to thank Fran\c cois Dubin for initiating this work and for numerous  insights. 
\end{acknowledgements}

\appendix
%\numberwithin{equation}{section}

\section{Resolution of Eq.~(\ref{66})\label{app:A}} 

The  analytical solution of Eq.~(\ref{66}) follows from the Cardan's trick \cite{nickalls1993}: we write $\mathcal{E}'$ as $U+ a/U$ and we choose $a$ such that the resulting equation 
\begin{equation}\label{Cardan}
0 = U^3 + \frac{a^3}{U^3} + \left( U + \frac{a}{U} \right) \left( 3a + z_2 \right) - z_3
\end{equation} has no term in $U$ and $1/U$. This leads to $a=-z_2 / 3$. The above equation then reduces to $0=U^6-z_3 U^3+a^3$. Its solutions simply read    
\begin{equation}
U_{\pm}^3 =  \frac{z_3}{2} \pm \sqrt{ \left(\frac{z_3}{2} \right)^2 + \left( \frac{z_2}{3} \right)^3 }\,, 
\end{equation}
which leads to
\begin{eqnarray}\label{A3}
U_{m,\pm} = e^{i m 2\pi/3}\left[ \frac{z_3}{2} \pm \sqrt{ \left( \frac{z_3}{2} \right)^2 + \left( \frac{z_2}{3} \right)^3  }\right]^{1/3}  \,
\end{eqnarray}
for $m=(0,\pm1)$. 
The resulting solutions of Eq.~(\ref{66}) are given by  
\begin{eqnarray}
\mathcal{E}'_{m,\pm}  &=& U_{m,\pm}  - \frac{z_2 / 3}{U_{m,\pm} } \,.\label{Ejn} 
\end{eqnarray}

Note that Eq. (\ref{A3}) seems to give six values while Eq.~(\ref{66}) is a cubic equation with only three solutions. To fix this problem, we note that $U_{m,+}  U_{-m,-}=-z_2/3$; so, the solutions with a $+$ and $-$ sign in Eq.~(\ref{A3}) are related by $\mathcal{E}'_{m,+} =\mathcal{E}'_{-m,-}$. Consequently, the three solutions of Eq.~(\ref{66}) can be taken either as the three values of $\mathcal{E}'_{m,+}$ or the three values of $\mathcal{E}'_{-m,-}$.

 %add titles in biblio: put in bib file
% @CONTROL{REVTEX41Control}
%@CONTROL{apsrev41Control,author="00",editor="1",pages="1",title="0",year="0"}
% put in tex file
%	\nocite{apsrev41Control}

%	\bibliographystyle{apsrev4-1} % Tell bibtex which bibliography style to use
%	\bibliography{../../main.bib}


\end{document}